\documentclass[%
 reprint, amsmath,amssymb, nofootinbib, aps, onecolumn
]{revtex4-1}

\usepackage{amsmath,amssymb,amsthm, graphicx, mathrsfs, pgf,tikz,marvosym,fontawesome,slashed, mathtools,array,bbm,yfonts, adjustbox, adforn, fourier-orns, relsize,stmaryrd,simplewick,youngtab,ytableau,youngtab, epigraph, url, bclogo,arydshln,multirow,dsfont, lipsum, tensor,feynmf, wrapfig, natbib, enumitem,subfig,verbatim}
\usepackage{physics}
\usepackage[left=20mm,right=20mm,top=35mm,bottom=35mm,columnsep=15pt]{geometry} 
\usepackage{placeins}
\usepackage[T1]{fontenc}
\usepackage[utf8]{inputenc}
\usepackage[nottoc]{tocbibind}
\usetikzlibrary{arrows}
\usetikzlibrary{positioning}
\usepackage[english]{babel}

\usepackage[uppercase, tiny, center]{titlesec}
\usepackage{cancel}

\definecolor{r}{cmyk}{1,.50,0,.20} 
\usepackage[colorlinks=true, linkcolor=r, urlcolor=r, linktocpage=true, citecolor=r]{hyperref}

\numberwithin{equation}{section}

\renewcommand{\d}{\textrm{d}}

\setcitestyle{square,comma}
\begin{document}

\title{\textbf{Cosmic eggs to relax the cosmological constant}}

\author{Thomas Hertog}
\email{thomas.hertog@kuleuven.be}
\affiliation{Institute  for  Theoretical Physics,  KU Leuven, Celestijnenlaan  200D, B-3001  Leuven, Belgium.}

\author{Rob Tielemans}
\email{rob.tielemans@kuleuven.be}
\affiliation{Institute  for  Theoretical Physics,  KU Leuven, Celestijnenlaan  200D, B-3001  Leuven, Belgium.}

\author{Thomas Van Riet}
\email{thomas.vanriet@kuleuven.be}
\affiliation{Institute  for  Theoretical Physics,  KU Leuven, Celestijnenlaan  200D, B-3001  Leuven, Belgium \\ $\&$  Institutionen f\"{o}r Fysik och Astronomi, Box 803, SE-751 08 Uppsala, Sweden}

\begin{abstract}
In theories with extra dimensions, the cosmological hierarchy problem can be thought of as the unnaturally large radius of the observable universe in Kaluza-Klein units. We sketch a dynamical mechanism that relaxes this. 
In the early universe scenario we propose, three large spatial dimensions arise through tunneling from a `cosmic egg', an effectively one-dimensional configuration with all spatial dimensions compact and of comparable, small size. If the string landscape is dominated by low-dimensional compactifications, cosmic eggs would be natural initial conditions for cosmology. A quantum cosmological treatment of a toy model egg predicts that, in a variant of the Hartle-Hawking state, cosmic eggs break to form higher dimensional universes with a small, but positive cosmological constant or quintessence energy. Hence cosmic egg cosmology yields a scenario in which the seemingly unnaturally small observed value of the vacuum energy can arise from natural initial conditions.  
\end{abstract}

\maketitle
\thispagestyle{empty}

\newpage

\section{Introduction}

The cosmological constant (cc) problem is one of the most vexing and longstanding conceptual problems in theoretical physics --for a review see e.g. \cite{Weinberg:1988cp,Martin:2012bt, Burgess:2013ara, Padilla:2015aaa}. It is therefore reasonable to anticipate that a better understanding of the origin of the observed small value of the vacuum energy will yield new insights in the fundamental physical laws and their status in a cosmological context. The cc problem arises when one couples quantum field theory (QFT) to gravity. In modern parlance, the problem follows directly from the fact that the vacuum energy is a relevant operator of dimension four. This means it is highly sensitive to the exact details of the UV physics which leave the low-energy effective field theory (EFT) otherwise unaffected. In particular any particle species with mass $m$ in the UV contributes, at one-loop, a term of the form\footnote{We note, however, that the calculation of the quantum contributions to the cc in QFT coupled to classical gravity isn't entirely unambiguous due to radiative instability. Naturalness arguments are therefore inherently qualitative.}
\begin{equation}
\Lambda M_p^2 \supset m^4 \log(\frac{m^2}{\mu^2})\,,
\end{equation} 
where $\mu$ is an RG scale, the precise meaning of which in this context being a matter of debate --see e.g. \cite{Shapiro2000,Foot2007,Shapiro2009,Ward2009,Hamber2013,Ward2014,Kohri2016}. The dominant contribution to $\Lambda$ comes from the heaviest particles with masses near the cut-off of the theory. Hence any fine-tuning of the bare cc to cancel contributions from particles in loops depends sensitively on the UV. 

In the context of the Standard Model EFT coupled to classical gravity, the natural value of the cosmological constant would therefore be $\Lambda \approx M^2_{\rm NP}$, where $M_{\rm NP}$ is the cut-off scale at which new physics (`NP') beyond the Standard Model enters. Conservatively one can take $M_{\rm NP} =M_{p}$ but in general $M_{\rm NP}$ can of course be lower. A straightforward formulation of the cc problem is that the observed dark energy density\footnote{The observed dark energy can be dynamical as e.g. in quintessence \cite{Caldwell:1997ii, Tsujikawa:2013fta}, but this hardly changes the crux of the problem.} $M_p^2\Lambda \approx (10^{-3} \rm eV)^4$ is extremely --unnaturally-- small in units of $M_{\rm NP}$:
\begin{equation}
\frac{\Lambda}{M_{\rm NP}^2}\ll 1\,.
\end{equation}
In terms of the length scales $L_{\Lambda}=|\Lambda|^{-1/2}$ and  $L_{\rm NP} = M_{\rm NP}^{-1}$ the cc problem amounts to the observation that 
\begin{equation}\label{ccprob2}
\frac{L_{\Lambda}}{L_{\rm NP}}\gg 1\,.
\end{equation}
Put differently, the observed Hubble scale $L_\Lambda$ is very much larger than its natural size $\sim L_{\rm NP}$? The supreme difficulty of the cc problem stems precisely from the fact that, at least within our current theoretical framework, it intertwines the largest and the smallest scales in physics. 

This strongly indicates that the cc problem can only be properly analysed --and hopefully ultimately resolved-- in the context of a UV-complete description of gravity such as string theory. It is a striking --yet underappreciated-- property of string theory indeed that the theory enables a precise calculation of the vacuum energy in certain controlled corners, at weak coupling and low energies. Celebrated examples include holographic backgrounds such as $\text{AdS}_5 \times S^5$ in IIB string theory, or $\text{AdS}_4\times S^7$ in 11-dimensional supergravity. At weak coupling and small curvature these are trusted backgrounds of string/M-theory, with a negative vacuum energy that does not receive significant quantum corrections, despite the presence of an infinite tower of light and heavy modes\footnote{At first sight one might think that these examples rely on supersymmetry to cancel quantum corrections. However, the same reasoning applies to non-supersymmetric vacua. In any controlled vacuum with weak curvature and small string coupling, all corrections must be sub-leading to the classical (low-energy) contributions to $\Lambda$, contrary to what one would expect on the basis of standard QFT-based reasoning.}. 

Hence string theory brings to the table genuinely novel ingredients which bear directly on the cc problem. Here we concentrate on two such elements, namely the possibility to decompactify dimensions and the shear abundance of string theory vacua. We then use quantum cosmology to combine these into a new toy model cosmological scenario which, we argue, offers an interesting new take on the cc problem.

First, consider the number of dimensions. In string phenomenology one usually starts with ten  dimensions of which six are compactified to arrive at low-energy solutions (vacua) of the form,
\begin{equation}\label{ansatz}
\d s^2_{10} =\d s^2_4 + \d s^2_6\,,
\end{equation}
with three non-compact spatial dimensions. 
From a cosmological viewpoint, an ansatz of this form amounts to a highly contrived starting point. After all how did the universe get there? Specifically, a genuinely four-dimensional compactification requires the following scale separation,
\begin{equation}\label{ccprob3}
\frac{L_{\Lambda}}{L_{\rm KK}} \gg 1\,, 
\end{equation}
where $L_{\rm KK}$ is the Kaluza-Klein (KK) scale associated to $\d s_6$ and $L_{\Lambda}$ is the, possibly time-dependent, Hubble scale on $\d s_4$. However this scale separation already implies a cc problem because $L_{\rm KK}$ is a `new physics' scale and hence \eqref{ccprob2} is identical to \eqref{ccprob3} \cite{Gautason:2015tig}. Four-dimensional compactifications obeying \eqref{ccprob3}, therefore, have a cc problem built in.
Relatedly, equation \eqref{ccprob3} means that standard four-dimensional QFT arguments apply only when the vacuum already has an unnaturally small cc. Hence to address the cc problem properly one must go beyond four-dimensional QFT and probe the origin of such scale separated configurations.

This hierarchy problem is equally clear from a ten-dimensional viewpoint. The ten-dimensional geometry \eqref{ansatz} carries two vastly different scales --the two most extreme scales in nature indeed. The corresponding fine-tuning is reflected in the difficulties to construct (meta)stable vacua exhibiting scale separation\footnote{In the limit of ten-dimensional supergravity, a no-go theorem for obeying (\ref{ccprob3}) \cite{Gautason:2015tig} extends the no-go theorem in \cite{Maldacena:2000mw}. In ten-dimensional supergravity with orientifolds, some constructions do yield precise and fully moduli-stabilised vacua with scale separation (see e.g. \cite{DeWolfe:2005uu, Farakos:2020phe}). Most compactifications with orientifolds, however, are of the no-scale type \cite{Dasgupta:1999ss, Giddings:2001yu}, meaning they feature Minkowksi vacua with some modes not stabilised. Stabilising these using quantum corrections, as in KKLT \cite{Kachru:2003aw} or LVS \cite{Balasubramanian:2005zx}, can give solutions with scale separation (\ref{ccprob3}), but never to arbitrary precision. Indeed the consistency of KKLT vacua is still being debated \cite{Polchinski:2015bea, Danielsson:2018ztv, Cicoli:2018kdo, Kachru:2018aqn,Gao:2020xqh}.} or, for that matter, gently rolling quintessence backgrounds \cite{Hebecker:2019csg}. However the latter can at least evade the swampland constraints \cite{Agrawal:2018own}. 

These difficulties motivate exploring different `lines of attack' on the cc problem. In this spirit, we consider a scenario in which all spatial dimensions are initially compactified and of comparable, small size. Such configurations can be thought of as effectively one-dimensional vacua or `cosmic eggs'. The absence of any scale separation means that cosmic eggs do not suffer from a cc problem or hierarchy problem. These eggs can be metastable and decay via tunneling to a configuration with  large expanding dimensions\footnote{Earlier cosmologies involving dynamical decompactification include string gas cosmology \cite{Brandenberger:1988aj} (see also \cite{Greene:2009gp, Greene:2012sa, Kim:2011cr}) and other scenarios like \cite{Heckman:2018mxl, Heckman:2019dsj}.}. Hence they form a natural starting point for cosmological considerations. We put forward a toy model cosmic egg cosmology in which an effectively four-dimensional universe emerges from the 
breaking and subsequent decompactification of a cosmic egg. Our model involves besides a cosmological constant\footnote{For simplicity we work with an effective four-dimensional cosmological constant but our results can be readily generalized to quintessence models.} also an axion which sets the overall size of the egg. 

It should be noted that the shear abundance of string vacua also means that there is not a unique cosmic egg. On the contrary, even the most ardent advocates of the swampland program would agree, we believe, that string theory encompasses a `landscape' of possible low-energy laws, albeit one with a set of interesting theoretical patterns that constrain its phenomenology. On general grounds one expects that vacua with a lower number of large dimensions are statistically hugely favored since there are more compact manifolds, more cycles to wrap branes and fluxes, etc. The `landscape' of higher dimensional vacua may well be a set of measure zero in the landscape of all vacua.

This proliferation of cosmic eggs means that a cosmic egg cosmology is neither complete nor predictive without specifying a state on its configuration space that defines a notion of typicality. We therefore embed cosmic egg cosmology in quantum cosmology, where the wave function of the universe provides a relative weighting of different egg-born cosmological histories from which a measure for observations can be obtained (see e.g. \cite{Hawking:2006ur,Hartle:2010dq,Hertog:2013mra}). Specifically we consider a toy model landscape consisting of a collection of cosmic eggs with different values of axion flux and $\Lambda$ in a variant of the Hartle-Hawking state. In this context we show that the model predicts the decay of a cosmic egg into an expanding, higher dimensional universe with a small positive cosmological constant. The latter is therefore seen to arise from natural initial conditions.


\section{Cosmic eggs}

To implement our scenario we are led to consider compactifications of string theory down to one dimension --time-- that serve as possible initial conditions for cosmic egg cosmology. The landscape of one-dimensional string theory vacua is relatively unexplored. Some preliminary work has been done in \cite{Haupt:2008nu, Haupt:2009hw, Heckman:2018mxl, Heckman:2019dsj}. We will not attempt to find new compactifications to one dimension (1D) but simply work with a toy model landscape that admits configurations that enable us to implement a cosmic egg cosmological scenario. 

Obviously we are interested in models that have 1D backgrounds that are metastable. This implies they probably should break supersymmetry. However, there are no 1D vacua in a strict sense once supersymmetry is broken \cite{Heckman:2018mxl,ArkaniHamed:2007gg}. This means one should not expect cosmic eggs to be perfectly static configurations. Second, we are interested in 1D backgrounds that can potentially tunnel to 4D universes\footnote{We comment below on cosmic egg cosmologies with a different number of large dimensions.}. A convenient way to proceed is to dimensionally reduce 4D EFTs. Consider the following action\footnote{From here onward, we work in Planck units $M_p=1$.}
\begin{equation}\label{action1}
S = \int \d^4x \sqrt{-g}\left(\frac{1}{2}\mathcal{R} -\Lambda- \frac{1}{2}H_3^2 \right)\,,
\end{equation}
where $H_3$ is an axion three-form field strength and $\Lambda$ is a four-dimensional cc, not necessarily small. We also note that adding more axions or two-form fluxes or even a rolling scalar potential does not significantly change the qualitatively analysis we wish to pursue here.

A dimensional reduction of this system over the three spatial dimensions amounts effectively to adopting a FLRW-like ansatz,
\begin{equation} \label{eq: metric Ansatz}
    \d s_4^2= -\frac{N^2(t)}{x(t)}\d t^2+x(t)\d s_3^2\,,
\end{equation}
where $N$ is a lapse function, $x$  acting as the volume modulus of a compact three-dimensional space with line element $ds^2_3$, which we take to be Einstein. 
The spatial curvature can be normalised as usual, in FLRW language $k=-1,0,+1$ with all three values consistent with compactness. 
The effective action for the volume-modulus is of the form
\begin{equation}
  S =-\frac{3}{2}\text{Vol}_3\int\d t\:N\left(\frac{1}{2N^2}\dot{x}^2-U(x)\right), \label{eq4: 1D vacuum}
\end{equation}
with
\begin{equation}
U(x) = -\frac{q^2}{3x^2}+2k-\frac{2}{3} \Lambda x. \label{eq: potential}
\end{equation}
where $q$ is the axion-flux.\footnote{To restore units, the length dimensions of the parameters are $[N]=L,[x]=L^2, [Q]=L^\alpha,[q]=L$ and $[\Lambda]=L^{-2}$.}

Gravity in one dimension is non-dynamical and governed by a Hamiltonian constraint --the first Friedmann equation-- that is enforced by the equation of motion for the lapse function $N$. Hence a 1D, time independent, (meta)stable vacuum only exists when $U$ has a minimum where it \emph{exactly} vanishes. It is hard, if not impossible, to achieve this with broken supersymmetry since quantum corrections are unlikely to sum exactly to zero. We do not want to trade one fine-tuning problem for another. However, for our purposes, it suffices for the potential $U$ to have a `pocket' region which can trap the scale factor. This would (in a weak sense) stabilize the volume modulus and create the possibility of an effective one-dimensional cosmological phase. 

\begin{figure}[t]
    \centering
    \includegraphics{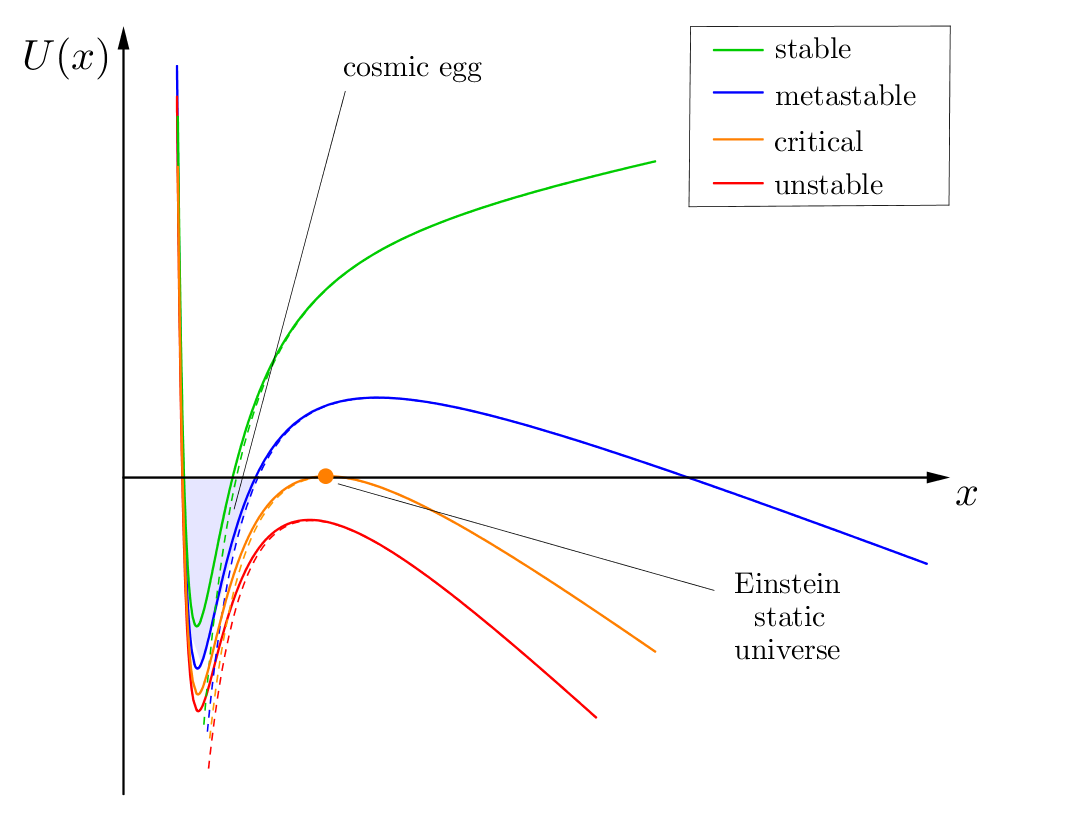}
    \caption{The effective potential $U$ for the volume modulus $x$ for a given axion-flux $q$, positive spatial curvature $k=+1$ and different values of the 4D cosmological constant $\Lambda$. The solid (dashed) lines represent the potential with (without) putative repulsive quantum corrections at small $x$ added.}
    \label{fig: cosmic eggs}
\end{figure}

If the spatial curvature is positive the potential has a maximum at zero energy when
\begin{equation}\label{ESU}
q^2 = \frac{8k}{\Lambda^2}\,.
\end{equation}
This corresponds to the unstable Einstein static Universe (cf. Fig \ref{fig: cosmic eggs}) and is not a candidate cosmic egg\footnote{Moreover since $q$ is quantised, the required tuning for this solution to exist may not be possible.}. The eggs, if they exist, are to be found at yet smaller values of the volume-modulus, where the potential becomes negative.

Now, one expects quantum corrections to modify the effective potential \eqref{eq: potential} at small $x$, adding terms involving higher inverse powers of $x$. For example, higher derivative terms come with extra inverse metrics leading to terms of the form $\sim x^{-2}$ or higher. An $F^4$ higher derivative correction in the Maxwell field, for instance, would introduce a term proportional to $p^4 x^{-5/2}$, with $p$ the magnetic flux. To compute whether $U$ will ultimately be repulsive or attractive at very small volume requires full control over the quantum gravity theory which is not within reach. Interestingly, however, models of loop quantum gravity suggest a repulsive behavior \cite{Bojowald:2018gdt}.  Figure \ref{fig: cosmic eggs} shows a few examples of an effective potential with a repulsive term at small $x$ added, for different values of the 4d cc $\Lambda$, and with $k=+1$.

The essential ingredient of our toy model that makes a cosmic egg scenario possible is the existence of a potential well at small volume in which the scale factor can be trapped and oscillate until it quantum mechanically tunnels through the barrier. As anticipated any cosmic egg is not expected to be static but rather exhibit a `breathing' volume modulus \cite{ArkaniHamed:2007gg}.  

Whether in our model an appropriate potential well exists, depends on three properties. First, the spatial curvature should be positive ($k=+1$). Second, $\Lambda$ should be bounded from above by the value that roughly corresponds to the Einstein static universe (\ref{ESU}), 
\begin{equation}
\Lambda \lesssim \Lambda_{\rm cr}\equiv  \frac{\sqrt{8}}{q}\,.\label{eq: cc critical}
\end{equation}
Third, the cc must be positive in order for the cosmic egg to be metastable. Thus within the range
\begin{equation}
    0<\Lambda < \Lambda_{\rm cr}.\label{eq: range cc}
\end{equation}
breathing cosmic eggs can emerge as novel, potentially natural early universe configurations, with all spatial dimensions of similar size. 

We next turn to the quantum cosmological dynamics of this model to identify the breathing, classical cosmic egg configurations and their evolution\footnote{A quantum cosmology analysis of a model that shares similarities with ours was presented in \cite{Bouhmadi-Lopez:2017sgq}. The context and motivation are, however, somewhat different.}.


\section{Breaking cosmic eggs}

Within the above range of values of $\Lambda$, cosmic eggs are metastable configurations. The eggs eventually `break' by tunneling through the potential barrier, causing the three spatial dimensions to decompactify and grow exponentially. This transition corresponds to the birth of a four-dimensional expanding universe. 

Quantum cosmology provides a unified treatment of cosmic egg cosmology, from the quantum formation of the classical oscillatory egg phase, to its breaking and the subsequent expansion of the universe. Furthermore in a landscape context, the wave function of the universe yields a relative weighting of different egg-born cosmological histories. Here we concentrate on a mini-landscape that is a collection of 1D vacua (eggs) with different values of the 4D cc $\Lambda$ and axion-flux. 

The compactified egg phase is ideally suited to be treated in a minisuperspace approximation, with the overall volume of the three compact spatial dimensions as the only remaining light degree of freedom. Of course new degrees of freedom will become relevant when the expansion gets underway. However our toy model does not include these, and hence does not describe a realistic universe. We are primarily interested here in modeling the earliest stages of evolution, the egg and tunneling phases, and to determine the relative weighting of expanding universes emerging from it. For these purposes it is reasonable to assume the minisuperspace approximation should apply. Upon quantization, \eqref{eq4: 1D vacuum} leads to the minisuperspace Wheeler-DeWitt equation,
\begin{equation}
    \left[-\frac{1}{2}\frac{\d^2}{\d x^2}+U(x)\right]\Psi(x) = 0, \label{eq: WdW}
\end{equation}
where $U$ is the potential \eqref{eq: potential} \textit{plus} some quantum corrections modifying its behaviour at small $x$ as argued before. 

It remains to specify the minisuperspace wave function.  In compactifications where $U(x)$ is repulsive as $x \to 0$, the wave function must definitely decay and eventually tend to zero in the small volume limit. This is because for a repulsive potential, the problem resembles that of a particle with an infinite wall in quantum mechanics. This choice of boundary condition is akin to the original Hartle-Hawking proposal \cite{HartleHawking}. It embodies its motivation that the big bang is a genuine beginning in a physical sense.

In compactifications where $U(x)$ is attractive as $x \to 0$, it would seem that one has a choice of boundary conditions on the wave function at $x=0$. One can either require the wave function to vanish at the origin, or one can impose boundary conditions for which the wave function continues to rise towards the origin. The former case is again akin to the original Hartle-Hawking proposal. We will see below that the no-boundary condition on the wave function acts as an effective quantum barrier at small $x$, giving rise to an effective pocket in which the wave function oscillates and describes a classical breathing egg just as in the repulsive case above. In the latter case the wave function would diverge as $x \to 0$ and it is doubtful that it describes a physically meaningful, let alone approximately classical, egg phase. 

\begin{figure}[t]
    \centering
        {
        \begin{picture}(200,100)
            \put(0,0){\includegraphics[width=0.35\textwidth]{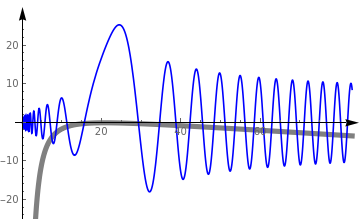}}
            \put(175,40){$x$}
            \put(-3,100){$\Psi$}
        \end{picture}\label{fig: wave function attractive}}
    \qquad
    {
        \begin{picture}(200,100)
            \put(0,0){\includegraphics[width=0.35\textwidth]{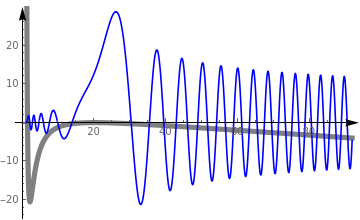}}
            \put(175,40){$x$}
            \put(-3,100){$\Psi$}
        \end{picture}\label{fig: wave function repulsive}}
    \caption{Numerical solution to the Wheeler-DeWitt equation \eqref{eq: WdW} with the Hartle-Hawking boundary condition $\Psi(0)=0$ for $q=29.4$ and $\Lambda=0.097$. The wave function is everywhere real and shown in blue, the potential is represented by the gray line. The left (right) panel shows the solution with an attractive (repulsive) potential in the small volume limit. Both models clearly exhibit a classical oscillatory egg state connected through quantum tunneling to an expanding four dimensional de Sitter-like universe.}
    \label{fig: wave function}
\end{figure}
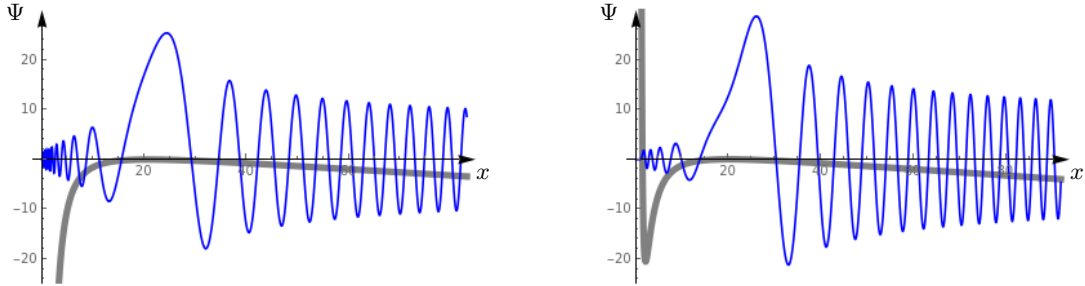

In both sets of models, we therefore require the wave function to decay to zero as $x \rightarrow 0$. Figure \ref{fig: wave function} shows an example of a numerical solution $\Psi (x)$ with `no-boundary' condition $\Psi (0)=0$, first in a model with an attractive small $x$ potential (such as \eqref{eq: potential}) and then with a repulsive `quantum-corrected' potential of the form,
\begin{equation}
    U(x) = \frac{Q^2}{x^\alpha}-\frac{q^2}{3x^2}+2k-\frac{2}{3} \Lambda x, \label{eq: potential WdW}
\end{equation}
with $\alpha>2$ and $k=+1$ as argued before (from here onwards, we will set $k=+1$). In both cases one can clearly identify the classical oscillatory egg state, the exponential behavior of the wave function under the barrier, and finally the classical expansion at large volume\footnote{See \cite{Hartle:2007gi,Hartle:2008ng} for an in depth discussion of the emergence of classical spacetime in quantum cosmology.}. The existence of a classical egg described by an oscillatory wave function requires $q^2>\frac{3}{8}M_p^2$ (cf. Appendix \ref{app: cosmic egg wavefunction}). Also, if $U$ is repulsive at small $x$, then $Q$ must be sufficiently small:\footnote{As $q^2$ only has a physical meaning, we write $q$ instead of $|q|$ to simplify notation.}
\begin{equation}
    Q^2 < \frac{2}{3\alpha}\left[\frac{1}{6}\left(1-\frac{2}{\alpha}\right)\right]^{\frac{\alpha}{2}-1}q^\alpha, \label{eq: bound Q}
\end{equation}
in order to have a region where the potential is negative. 

The steepness of the potential in the small volume limit causes the WKB approximation to break down there. This is rather different from the standard applications of Hartle-Hawking initial conditions in the context of inflation, involving gentle and finite, positive potentials in the small volume regime, in which the minisuperspace wave function can be expressed in terms of (the action of) regular semiclassical saddle points. In Appendix \ref{app: cosmic egg wavefunction} we present an approximate solution of the small $x$ wave function with Hartle-Hawking boundary conditions using a different approximation scheme.

On the other hand the WKB approximation becomes accurate for larger $x$. The general behavior of the WKB solution in the large volume $x\gg x_3$ regime reads,
\begin{equation}\label{WKB}
    \Psi(x) = \frac{1}{|2U(x)|^{1/4}}\left(A e^{iS(x,x_3)}+ Be^{-iS(x,x_3)}\right)\,,
\end{equation}
where $S(x,x_i)$ is defined as
\begin{equation}
    S(x,x_i)=\int_{x_i}^x\d x'\sqrt{|2U(x')|}.
\end{equation}
and $x_3$ is the large $x$ endpoint of the potential barrier (cf. Fig. \ref{fig: cosmic egg regions}). 

The standard WKB connection formulae determine the general form of the coefficients $A$ and $B$,
\begin{subequations}
\begin{align}
    & A = e^{S_{23}-i\frac{\pi}{4}}\mathcal{H} + e^{-S_{23}+i\frac{\pi}{4}}\mathcal{V}\\
    & B = e^{S_{23}+i\frac{\pi}{4}}\mathcal{H} + e^{-S_{23}-i\frac{\pi}{4}}\mathcal{V}
\end{align}\label{eq: connection formula1}%
\end{subequations}%
where $S_{23}\equiv S(x_2,x_3)$ and the coefficients $\mathcal{H}(Q,\alpha,q,\Lambda)$ and $\mathcal{V}(Q,\alpha, q,\Lambda)$ depend on the choice of boundary condition on $\Psi$. They are obtained by matching the small $x$ wave function to the WKB form \eqref{WKB}. We refer to Appendix \ref{app: cosmic egg wavefunction} for the details of this procedure. 

We have numerically verified and analytically substantiated in Appendix \ref{app: cosmic egg wavefunction} that both $\mathcal{H}$ and $\mathcal{V}$ are generically non-zero for Hartle-Hawking boundary conditions at $x=0$. Hence the intermediate tunneling dynamics in cosmic egg cosmology naturally produces a wave function in the large volume limit that involves a superposition of the typical Hartle-Hawking behavior $\sim e^{S_{23}}$, with a Vilenkin tail $\sim e^{-S_{23}}$. Substituting \eqref{eq: connection formula1} in \eqref{WKB} yields the large $x$ wave function
\begin{equation}
    \Psi(x) = \frac{2}{|2U(x)|^{1/4}}\left[e^{S_{23}}\mathcal{H}\cos\left(S(x,x_3)-\frac{\pi}{4}\right)+e^{-S_{23}}\mathcal{V}\cos\left(S(x,x_3)+\frac{\pi}{4}\right)\right]\qquad(x\gg x_3).
\end{equation}
The leading dependence on $q$ and $\Lambda$ is encoded in $S_{23}$. To understand this, note that $S_{23}$ is given by the surface area of the function $\sqrt{U}$ between the two turning points $x_2$ and $x_3$. With $H$ the maximum height of this and $\Delta x= |x_3-x_2|$ the distance between the turning points $x_2$ and $x_3$, the area can be shown to be somewhere between
\begin{equation}
\frac{H\Delta x}{2}< S_{23}<H\Delta x\,,    
\end{equation}
since $U(x)$ is concave. One then finds
\begin{equation}
 H \approx \sqrt{4 - 2\left(q^2\Lambda^2\right)^{1/3}}\,.
\end{equation}
An approximate expression for $\Delta x$, valid when $\Lambda$ is significantly lower than its critical value \eqref{eq: cc critical}, is given by
\begin{equation}
    \Delta x \approx \frac{3}{\Lambda} - \frac{q}{\sqrt{6}}\,.
\end{equation}
This yields the following estimate,
\begin{equation}
S_{23}\approx  \zeta \left(\frac{3}{\Lambda} - \frac{q}{\sqrt{6}}\right)\sqrt{4 - 2\left(q^2\Lambda^2\right)^{1/3}}\,, \label{eq: S23}
\end{equation}
with $1/2 < \zeta < 1$. In the limit $\Lambda\ll\Lambda_{\rm cr}$ this becomes
\begin{equation}
    S_{23} \approx \frac{3\zeta}{\Lambda}\left[2-\left(\frac{\Lambda}{\Lambda_{\rm cr}}\right)^{2/3}+\mathcal{O}\left(\frac{\Lambda}{\Lambda_{\rm cr}}\right)\right].
\end{equation}
Hence the wave function in this limit is approximately given by
\begin{equation}
    \Psi (x) \approx \frac{2\mathcal{H}e^{S_{23}}}{|2U(x)|^{1/4}}\cos\left(S(x,x_3)-\frac{\pi}{4}\right)\qquad(x\gg x_3)\,.
\end{equation}

Thus the wave function of the universe in the large volume regime in a cosmic egg scenario consists of a leading Hartle-Hawking term, given here, together with an exponentially suppressed contribution characteristic of the tunneling wave function. The mixing of both wave functions arises because the quantum dynamics of the egg cosmology involves a combination of Hartle-Hawking initial conditions to create the egg, and quantum tunneling from the egg to a large expanding universe.


\section{Discussion}

We have argued that the cc problem motivates the study of cosmological scenarios in which all spatial dimensions are initially of comparable, small size.  That is, cosmology encourages one to consider decompactification rather than compactification, as in string gas cosmology  \cite{Brandenberger:1988aj}. 

We have proposed a new toy model in this spirit in which a universe with three large spatial dimensions is seen to emerge in a two-step process. First, we conceive of a metastable, classical, fully compactified configuration in the Hartle-Hawking state. These effectively zero-dimensional cosmic egg configurations are neither de Sitter nor anti-de Sitter but breathing, with an oscillating volume modulus. Their size is set by a combination of the particle physics ingredients, including an axion, and the Hartle-Hawking initial conditions. The absence of scale separation means that cosmic eggs do not suffer from a cc problem and are thus plausibly natural early universe configurations\footnote{As a historical comment, we note that the cosmic egg configurations we consider, differ from Lema\^itres conception of `primeval atom' \cite{Lemaitre1} which he thought of as purely quantum, i.e. exhibiting no (classical) notion of space and time. Lema\^itre's primeval atom seems more akin to the Euclidean region of the conventional no-boundary saddle points.} 

When the cosmological constant term is positive but below the Einstein static value in the model, we have identified a decompactification channel in which the egg decays through tunneling into an expanding universe with three large spatial dimensions. In a landscape consisting of eggs with different parameters the characteristic Hartle-Hawking weighting then favors the nucleation of large universes with a low value of the four dimensional cc. This cast the cc problem in a new light: a seemingly unnaturally small cc is seen to arise from natural initial conditions. 

Whilst we have concentrated on the cc, much of our reasoning as well as our calculations carry over to quintessence models where the egg decays to a large expanding universe with a rolling value of the vacuum energy. We have also limited our attention to the creation of four-dimensional universes. However our mechanism would allow for the birth of higher dimensional universes too. So unless there is an obstruction in the string landscape, there is no reason to exclude these. That said, the nucleation of four rather than higher dimensional universes from eggs may be preferred on statistical grounds for the same reason that we have argued that eggs form natural initial conditions in the first place. To date no proposal in string theory exists to construct flux vacua with moduli stabilisation that feature scale separation in dimensions larger than four. 

The combination of Hartle-Hawking initial conditions in the small volume limit with a tunneling process at intermediate volumes, means that the minisuperspace wave function at large volume behaves as a superposition of the usual Hartle-Hawking and tunneling wave functions. Specifically, the wave function is dominated in the large volume limit by Hartle-Hawking saddles but receives an exponentially suppressed contribution characteristic of the tunneling wave function. The latter may be hard to avoid, unlike in scenarios considered previously in the literature, where classical expanding universes emerge directly from a quantum phase.\footnote{We refer to \cite{Feldbrugge:2017fcc,Feldbrugge:2017kzv,Feldbrugge:2017mbc,DiazDorronsoro:2017hti,DiazDorronsoro:2018wro,Janssen:2019sex,Halliwell:2018ejl} for a more detailed discussion of this.} 

Needless to say, the cosmic egg cosmologies we have considered aren't realistic. The sole purpose of the toy models we have proposed, is to exhibit a concrete mechanism to relax the cc. It will require much further work to embed this scenario in somewhat realistic models of the early universe. One possible route would appear to consider the breathing eggs as a pre-inflationary phase, with the decay process leading to a period of inflation that generates a large universe filled with matter and primordial perturbations. It is plausible that the general tendency of the Hartle-Hawking state towards a low vacuum energy survives in such more realistic egg-based cosmologies \cite{Hartle:2013oda}.

\begin{acknowledgments}
This work is supported by the C16/16/005 grant of the KU Leuven, the COST Action GWverse CA16104, and by the FWO Grant No. G092617N. TVR would like to thank the FWO-Vlaanderen and the KU Leuven for supporting his sabbatical research.
\end{acknowledgments}

\appendix
\numberwithin{equation}{section}
\newpage

\section{Cosmic egg wave function}\label{app: cosmic egg wavefunction}

\begin{figure}[t]
    \centering
    \includegraphics[width=0.9\textwidth]{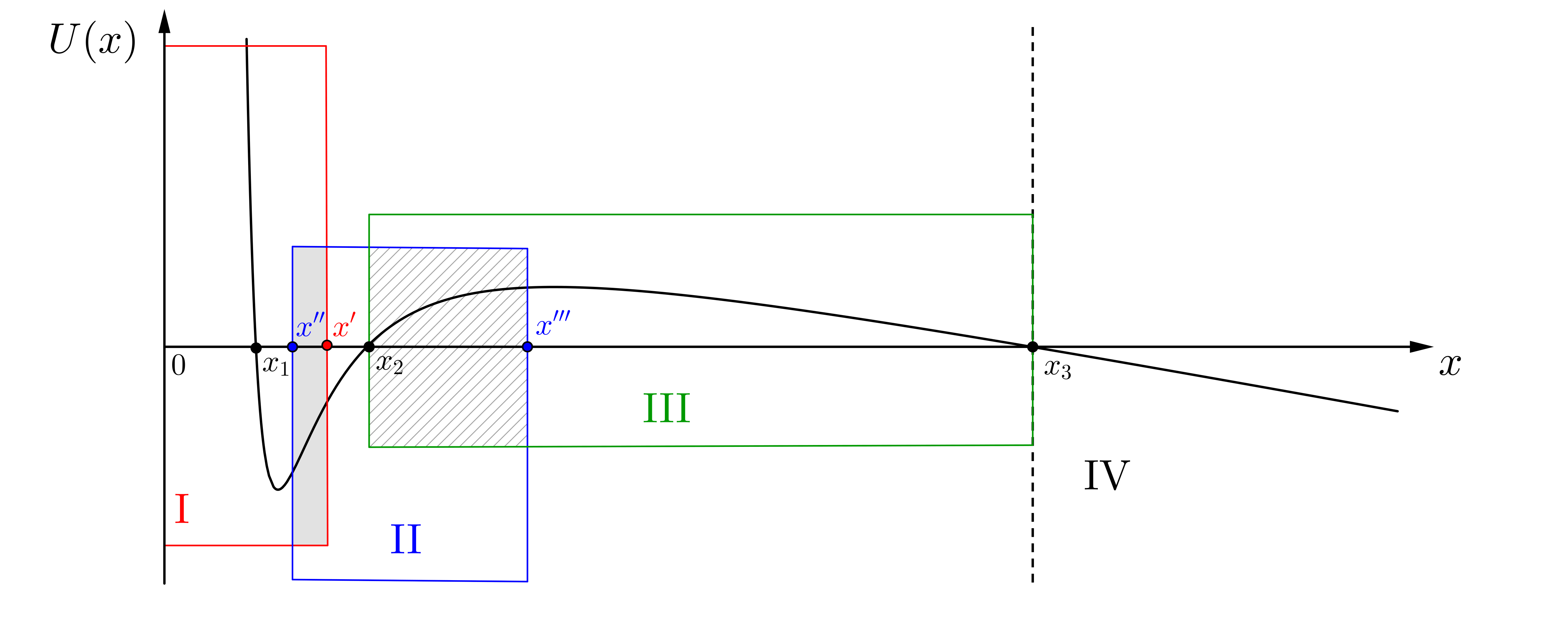}
    \caption{The different regions of the potential in which the Wheeler--DeWitt equation is solved. Solutions in region I, II and III are matched within the overlap (shaded and striped region) and connected to region IV with the WKB connection formula. }
    \label{fig: cosmic egg regions}
\end{figure}

In this Appendix we derive an approximate analytical solution of the wave function in the cosmic egg model beyond the WKB approximation which fails in the small volume regime. Specifically we first solve the Wheeler--DeWitt equation approximately in three different potential regions and then match the solutions across the regions and eventually with the WKB form at large $x$. Figure \ref{fig: cosmic egg regions} shows the subdivision in regions, for a potential of the form 
\begin{equation}
    U(x) = \frac{Q^2}{x^\alpha}-\frac{q^2}{3x^2}+2-\frac{2}{3}\Lambda x. \label{eqapp: potential}
\end{equation}%
It should be noted that the approximation solution derived below is valid for configurations that have $\Lambda\ll\Lambda_{\rm cr}$ and $x_1\ll x''\ll x'\ll x_2$ or, equivalently, 
\begin{equation}
    2^{\frac{1}{2}+\frac{1}{\alpha-2}}\ll\frac{x_2}{x_1},
\end{equation}
or
\begin{equation}
    Q^2\ll2\left(\frac{q}{\sqrt{12}}\right)^\alpha. \label{eq: domain validity}
\end{equation}
Note that \eqref{eq: domain validity} is stronger than the aforementioned bound on $Q$ in eq \eqref{eq: bound Q}, for all values of $\alpha>2$.

Region I is specified by $x\ll x'$ with
\begin{equation}
    x' = \frac{q}{2\sqrt{3}} \approx \frac{x_2}{\sqrt{2}}\,.
\end{equation}
In this region, the first two terms of the potential \eqref{eqapp: potential} dominate over the curvature and cc term, i.e.
\begin{equation}
    U(x) \approx \frac{Q^2}{x^\alpha}-\frac{q^2}{3x^2}\qquad\qquad\text{(region I)}\,.
\end{equation}
Consequently, the Wheeler--DeWitt equation is approximately solved by
\begin{equation}
    \Psi_{\rm I}(x) = a\:\sqrt{x}K_\nu\left(\frac{\tilde{Q}}{x^{\frac{\alpha}{2}-1}}\right)+b\:\sqrt{x}I_\nu\left(\frac{\tilde{Q}}{x^{\frac{\alpha}{2}-1}}\right), \label{eq: wave function small x}
\end{equation}
where $I_\nu$ and $K_\nu$ are the modified Bessel functions of the first and second order $\nu$ and
\begin{align}
    \nu = \frac{1}{\alpha-2}\sqrt{1-\frac{8}{3}q^2}\,, && \tilde{Q} = \frac{2\sqrt{2}}{\alpha-2}Q.
\end{align}
Hence $\Psi_{\rm I}$ only displays oscillations characteristic for a classical egg phase when 
\begin{equation}
    q^2>\frac{3}{8}. \label{eq: bound q}
\end{equation}
In what follows we will assume this lower bound on $q$. Also, in the limit $x\to0$, only the first term in \eqref{eq: wave function small x} remains regular and --in particular-- vanishes. The Hartle-Hawking wave function in region I is therefore given by eq. (\ref{eq: wave function small x}) with $a=1$ and $b=0$. 

In region II, any quantum effects $\sim Q^2/x^\alpha$ in the potential \eqref{eq: potential WdW} are damped out and the curvature term becomes relevant while the cc term remains negligible. This region roughly corresponds to $x''\ll x \ll x'''$ with 
\begin{subequations}
\begin{align}
    &x'' = \left(\frac{6 Q^2}{q^2}\right)^{\frac{1}{\alpha-2}} = 2^{\frac{1}{\alpha -2}}x_1\\
    &x''' = \frac{3}{2\Lambda} = \frac{x_3}{2}
\end{align}
\end{subequations}
The potential can be approximated by
\begin{equation}
    U(x) \approx -\frac{q^2}{3x^2}+2 \qquad\text{(region II)} \label{eq: potential II}
\end{equation}
and the wave function is then approximately given by
\begin{equation}
    \Psi_{\rm II}(x) = c\sqrt{x}K_{\mu}\left(2 x\right) + d\sqrt{x}M_\mu\left(2 x\right)\qquad\,,\label{eq: wavefunction II}
\end{equation}
where $M_\mu(x)$ is defined to be
\begin{equation}
    M_\mu(x)\equiv I_\mu(x)+I_{-\mu}(x).
\end{equation}
The choice of writing $K_\mu$ and $M_\mu$ in \eqref{eq: wavefunction II} simplifies the interpretation: if $c$ and $d$ are real, so is $\Psi_{\rm II}$. The order $\mu$ of the modified Bessel functions is again purely imaginary under the constraint \eqref{eq: bound q} and given by
\begin{equation}
    \mu = \left(\frac{\alpha}{2}-1\right)\nu\,. \label{eq: mu}
\end{equation}
The integration constants $c$ and $d$ are determined by the behaviour of the wave function in region I. This is done by a matching procedure in the overlap region between $x''$ and $x'$, the shaded area in Figure \ref{fig: cosmic egg regions}. For $ x''< x<x'$, the  behaviour of \eqref{eq: wave function small x}  is
\begin{align}
     \Psi_{\rm I}(x) \sim \tilde{Q}^\nu \frac{\Gamma(-\nu)}{2^{1+\nu}} x^{-\left(\frac{\alpha}{2}-1\right)\nu+\frac{1}{2}}+\frac{\tilde{Q}^{-\nu}}{2^{1-\nu}}\Gamma(\nu)x^{\left(\frac{\alpha}{2}-1\right)\nu+\frac{1}{2}}+\mathcal{O}\left(\left(\frac{x_1}{x}\right)^{\alpha-\frac{5}{2}\pm\left(\frac{\alpha}{2}-1\right)\nu}\right)\,,\label{eq: wavefunction I asy}
\end{align}
while for $x''<x<x'\ll x_2$, the wave function \eqref{eq: wavefunction II} takes the form
\begin{equation}
    \Psi_{\rm II}(x)\sim   x^{-\mu+\frac{1}{2}}\left(\frac{1}{2}\Gamma(\mu)c+\frac{d}{\Gamma(1-\mu)}\right)+x^{\mu+\frac{1}{2}}\left(\frac{1}{2}\Gamma(-\mu)c+\frac{d}{\Gamma(1+\mu)}\right)+\mathcal{O}\left(\left(\frac{x}{x_2}\right)^{\frac{5}{2}\pm\mu}\right)
\end{equation}
Hence the matching condition is
\begin{subequations}
\begin{align}
    &c =  \left(\frac{\tilde{Q}}{2}\right)^\nu\frac{\Gamma(1-\mu)\Gamma(-\nu)}{\Gamma(1-\mu)\Gamma(\mu)-\Gamma(-\mu)\Gamma(1+\mu)}+\rm c.c.\\
    &d = \frac{1}{2}\left(\frac{\tilde{Q}}{2}\right)^\nu\frac{\Gamma(1-\mu)\Gamma(1+\mu)\Gamma(-\mu)\Gamma(-\nu)}{\Gamma(-\mu)\Gamma(1+\mu)-\Gamma(1-\mu)\Gamma(\mu)}+\rm c.c.
\end{align}\label{eq: connection c d}%
\end{subequations}
where c.c. denotes complex conjugate. Hence it is obvious that $c$ and $d$ are real and, consequently, $\Psi_{\rm II}$ is real.

Within region II it is to be expected that the WKB approximation becomes increasingly accurate for larger $x$. We will therefore aim at giving WKB solutions in region III (overlapping with the right portion of region II) and IV. Region III is the underbarrier region between the turning points $x_2<x<x_3$ in which the wave function takes the WKB form
\begin{equation}
    \Psi_{\rm III}(x) = \frac{1}{|2U(x)|^{1/4}}\left(C e^{-S(x,x_2)}+D e^{S(x,x_2)}\right)\,, \label{eq: wavefunction III}
\end{equation}
where $S(x,x_i)$ is defined as
\begin{equation}
    S(x,x_i) = \int_{x_i}^x\d x'\sqrt{|2U(x')|}.
\end{equation}
The amplitudes $C$ and $D$ are to be matched with $c$ and $d$ that specify the behaviour of the wave function in the first part of the underbarrier regime (the striped area in Fig \ref{fig: cosmic egg regions}) where the cc term can still be neglected. In this limit, $x_2\ll x < x'''$, the asymptotic form of \eqref{eq: wavefunction II} is
\begin{equation}
    \Psi_{\rm II}(x)\sim \frac{1}{2}c\sqrt{\pi}e^{-2x}+\frac{d}{\sqrt{\pi}}e^{2x}+\mathcal{O}\left(e^{\pm 2x}\left(\frac{x_2}{x}\right)^{3/2}\right)
\end{equation}
while the WKB form \eqref{eq: wavefunction III} is approximately
\begin{equation}
    \Psi_{\rm III}(x)\sim \frac{C}{\sqrt{2}}e^{\frac{\pi}{\sqrt{6}}q}e^{-2 x}+\frac{D}{\sqrt{2}}e^{-\frac{\pi}{\sqrt{6}}q}e^{2 x}
\end{equation}
This is because in this limit $x_2\ll x < x'''$
\begin{equation}
    S(x,x_2)\approx 2 x\sqrt{1-\left(\frac{x_2}{x}\right)^2}-2 x_2\arctan\left(\sqrt{\left(\frac{x}{x_2}\right)^2-1}\right) \sim 2
    x - \frac{\pi}{\sqrt{6}}q+\mathcal{O}\left(\frac{x_2}{x}\right)\,
\end{equation}
where the potential \eqref{eq: potential II} has been inserted. Hence the amplitudes $C$ and $D$ are given by
\begin{subequations}
\begin{align}
   & C = \sqrt{\frac{\pi}{2}}e^{-\frac{\pi}{\sqrt{6}}q}c \\
   & D = \sqrt{\frac{2}{\pi}}e^{\frac{\pi}{\sqrt{6}}q}d
\end{align}\label{eq: connection C D}%
\end{subequations}%
To make the transition to region IV, $x_3<x$, one simply uses the standard WKB connection formulae. The wave function in this large $x$ regime is given by
\begin{equation}
    \Psi_{\rm IV}(x) = \frac{1}{|2U(x)|^{1/4}}\left(A e^{iS(x,x_3)}+B e^{-iS(x,x_3)}\right)\,, \label{eq: wavefunction IV}
\end{equation}
where 
\begin{subequations}
\begin{align}
    &A = De^{S_{23}-i\frac{\pi}{4}}+\frac{1}{2}Ce^{-S_{23}+i\frac{\pi}{4}}\\
    &B = De^{S_{23}+i\frac{\pi}{4}}+\frac{1}{2}Ce^{-S_{23}-i\frac{\pi}{4}}
\end{align}\label{eq: connection A B}%
\end{subequations}%
where $S_{23}\equiv S(x_2,x_3)$. As stated in the text, the connection formulae for this specific cosmic egg model are of the general form
\begin{subequations}
\begin{align}
    & A = e^{S_{23}-i\frac{\pi}{4}}\mathcal{H}(Q,\alpha,q) + e^{-S_{23}+i\frac{\pi}{4}}\mathcal{V}(Q,\alpha,q)\\
    & B = e^{S_{23}+i\frac{\pi}{4}}\mathcal{H}(Q,\alpha,q) + e^{-S_{23}-i\frac{\pi}{4}}\mathcal{V}(Q,\alpha,q)
\end{align}\label{eq: connection formula}%
\end{subequations}%
where $\mathcal{H}$ and $\mathcal{V}$ depend on the quantum contributions to the potential as well as the scalar flux. In particular, using equations \eqref{eq: connection c d}, \eqref{eq: connection C D} and \eqref{eq: connection A B}, $\mathcal{H}$ and $\mathcal{V}$ are explicitly given by
\begin{subequations}
\begin{align}
    &\mathcal{H} = \sqrt{\frac{2}{\pi}}e^{\frac{\pi}{\sqrt{6}}q}\Re\left[\left(\frac{\tilde{Q}}{2}\right)^\nu\frac{\Gamma(1-\mu)\Gamma(1+\mu)\Gamma(-\mu)\Gamma(-\nu)}{\Gamma(-\mu)\Gamma(1+\mu)-\Gamma(1-\mu)\Gamma(\mu)}\right] \\
    &\mathcal{V} =\sqrt{\frac{\pi}{2}}e^{-\frac{\pi}{\sqrt{6}}q}\Re\left[\left(\frac{\tilde{Q}}{2}\right)^\nu\frac{\Gamma(1-\mu)\Gamma(-\nu)}{\Gamma(1-\mu)\Gamma(\mu)-\Gamma(-\mu)\Gamma(1+\mu)}\right]
\end{align}
\end{subequations}

\bibliographystyle{toine}
\bibliography{refs}

\end{document}